\documentclass[twocolumn]{aastex631} 

\usepackage{gensymb}
\usepackage[normalem]{ulem}
\usepackage{hyperref}

\usepackage{graphicx} 

\received{--}
\revised{--}
\accepted{--}
\submitjournal{ApJL}

\newcommand{\WHL}{WHL0137$-$08}
\newcommand{\HST}{\emph{HST}}
\newcommand{\JWST}{\emph{JWST}}
\newcommand{\zphot}{\ensuremath{z_{\text{phot}}}}

\shorttitle{JWST Imaging of Earendel}
\shortauthors{Welch et al.}

\begin{document}

\title{JWST Imaging of Earendel, the Extremely Magnified Star at Redshift $z=6.2$}

\correspondingauthor{Brian Welch}
\email{bwelch7@jhu.edu}

\author[0000-0003-1815-0114]{Brian Welch}
\affiliation{Center for Astrophysical Sciences, Department of Physics and Astronomy, The Johns Hopkins University, 
3400 N Charles St. Baltimore, MD 21218, USA}

\author[0000-0001-7410-7669]{Dan Coe}
\affiliation{Space Telescope Science Institute (STScI), 3700 San Martin Drive, Baltimore, MD 21218, USA}
\affiliation{Association of Universities for Research in Astronomy (AURA) for the European Space Agency (ESA), STScI, Baltimore, MD, USA}
\affiliation{Center for Astrophysical Sciences, Department of Physics and Astronomy, The Johns Hopkins University, 
3400 N Charles St. Baltimore, MD 21218, USA}

\author[0000-0003-1096-2636]{Erik Zackrisson}
\affiliation{Observational Astrophysics, Department of Physics and Astronomy, Uppsala University, Box 516, SE-751 20 Uppsala, Sweden}

\author[0000-0001-9336-2825]{S.~E.~de~Mink}
\affiliation{Max-Planck-Institut für Astrophysik, Karl-Schwarzschild-Straße 1, 85741 Garching, Germany}
\affiliation{Anton Pannekoek Institute for Astronomy and GRAPPA, University of Amsterdam, NL-1090 GE Amsterdam, The Netherlands}

\author[0000-0002-5269-6527]{Swara Ravindranath}
\affiliation{Space Telescope Science Institute (STScI), 3700 San Martin Drive, Baltimore, MD 21218, USA}

\author[0000-0003-2861-3995]{Jay Anderson}
\affiliation{Space Telescope Science Institute (STScI), 3700 San Martin Drive, Baltimore, MD 21218, USA}

\author[0000-0003-2680-005X]{Gabriel Brammer}
\affiliation{Cosmic Dawn Center (DAWN), Copenhagen, Denmark}
\affiliation{Niels Bohr Institute, University of Copenhagen, Jagtvej 128, Copenhagen, Denmark}

\author[0000-0002-7908-9284]{Larry Bradley}
\affiliation{Space Telescope Science Institute (STScI), 3700 San Martin Drive, Baltimore, MD 21218, USA}

\author[0000-0002-4168-239X]{Jinmi Yoon}
\affiliation{Space Telescope Science Institute (STScI), 3700 San Martin Drive,  Baltimore, MD 21218, USA}
\affiliation{Joint Institute for Nuclear Astrophysics - Center for the Evolution of the Elements, USA}

\author[0000-0003-3142-997X]{Patrick Kelly}
\affiliation{School of Physics and Astronomy, University of Minnesota, 116 Church Street SE, Minneapolis, MN 55455, USA}

\author[0000-0001-9065-3926]{Jose M. Diego}
\affiliation{Instituto de F\'isica de Cantabria (CSIC-UC). Avda. Los Castros s/n. 39005 Santander, Spain}

\author[0000-0001-8156-6281]{Rogier Windhorst}
\affiliation{School of Earth and Space Exploration, Arizona State University, Tempe, AZ 85287, USA}

\author[0000-0002-0350-4488]{Adi Zitrin}
\affiliation{Physics Department, Ben-Gurion University of the Negev, P.O. Box 653, Be'er-Sheva 84105, Israel}

\author[0000-0001-7399-2854]{Paola Dimauro}
\affiliation{INAF - Osservatorio Astronomico di Roma, via di Frascati 33, 00078 Monte Porzio Catone, Italy}

\author[0000-0002-6090-2853]{Yolanda Jim\'enez-Teja}
\affiliation{Instituto de Astrof\'isica de Andaluc\'ia, Glorieta de la Astronom\'ia s/n, 18008 Granada, Spain}
\affiliation{Observatório Nacional - MCTI (ON), Rua Gal. José Cristino 77, São Cristóvão, 20921-400, Rio de Janeiro, Brazil}

\author[0000-0002-5258-8761]{Abdurro'uf}
\affiliation{Institute of Astronomy and Astrophysics, Academia Sinica, 11F of AS/NTU Astronomy-Mathematics Building, No.1, Sec. 4, Roosevelt Road, Taipei 10617, Taiwan, R.O.C.}
\affiliation{Center for Astrophysical Sciences, Department of Physics and Astronomy, The Johns Hopkins University, 
3400 N Charles St. Baltimore, MD 21218, USA}
\affiliation{Space Telescope Science Institute (STScI), 
3700 San Martin Drive, 
Baltimore, MD 21218, USA}

\author[0000-0001-6342-9662]{Mario Nonino}
\affiliation{INAF-Trieste Astronomical Observatory, Via Bazzoni 2, 34124, Trieste, Italy}


\author[0000-0003-3108-9039]{Ana Acebron}
\affiliation{Dipartimento di Fisica, Universit\`a degli Studi di Milano, Via Celoria 16, I-20133 Milano, Italy}
\affiliation{INAF - IASF Milano, via A. Corti 12, I-20133 Milano, Italy}

\author[0000-0002-8144-9285]{Felipe Andrade-Santos}
\affiliation{Department of Liberal Arts and Sciences, Berklee College of Music, 7 Haviland Street, Boston, MA 02215, USA}
\affiliation{Center for Astrophysics \text{\textbar} Harvard \& Smithsonian, 60 Garden Street, Cambridge, MA 02138, USA}

\author[0000-0001-9364-5577]{Roberto J. Avila}
\affiliation{Space Telescope Science Institute (STScI),  3700 San Martin Drive,  Baltimore, MD 21218, USA}

\author[0000-0003-1074-4807]{Matthew B. Bayliss}
\affiliation{Department of Physics, University of Cincinnati, Cincinnati, OH 45221, USA}

\author{Alex Ben\'itez}
\affiliation{King's College London, University of London, Strand, London WC2R 2LS, UK}

\author[0000-0002-8785-8979]{Tom Broadhurst}
\affiliation{Department of Theoretical Physics, University of the Basque Country UPV/EHU, Bilbao, Spain}
\affiliation{Donostia International Physics Center (DIPC), 20018 Donostia, Spain}
\affiliation{IKERBASQUE, Basque Foundation for Science, Bilbao, Spain}

\author[0000-0003-0883-2226]{Rachana Bhatawdekar}
\affiliation{European Space Agency, ESA/ESTEC, Keplerlaan 1, 2201 AZ Noordwijk, NL}

\author[0000-0001-5984-0395]{Maru{\v s}a Brada{\v c}}
\affiliation{University of Ljubljana, Department of Mathematics and Physics, Jadranska ulica 19, SI-1000 Ljubljana, Slovenia}
\affiliation{Department of Physics and Astronomy, University of California Davis, 1 Shields Avenue, Davis, CA 95616, USA}

\author[0000-0001-6052-3274]{Gabriel B. Caminha}
\affiliation{Max-Planck-Institut für Astrophysik, Karl-Schwarzschild-Straße 1, 85741 Garching, Germany}

\author[0000-0003-1060-0723]{Wenlei Chen}
\affiliation{School of Physics and Astronomy, University of Minnesota, 116 Church Street SE, Minneapolis, MN 55455, USA}


\author[0000-0002-1722-6343]{Jan Eldridge}
\affiliation{Department of Physics, University of Auckland, Private Bag 92019, Auckland, New Zealand}

\author[0000-0002-5794-4286]{Ebraheem Farag}
\affiliation{Arizona State University, Tempe, AZ 85287, USA}


\author[0000-0001-5097-6755]{Michael Florian}
\affiliation{Department of Astronomy, Steward Observatory, University of Arizona, 933 North Cherry Avenue, Tucson, AZ 85721, USA}

\author[0000-0003-1625-8009]{Brenda Frye}
\affiliation{Department of Astronomy, Steward Observatory, University of Arizona, 933 North Cherry Avenue, Tucson, AZ 85721, USA}

\author[0000-0001-7201-5066]{Seiji Fujimoto}
\affiliation{Cosmic Dawn Center (DAWN), Copenhagen, Denmark}
\affiliation{Niels Bohr Institute, University of Copenhagen, Jagtvej 128, Copenhagen, Denmark}

\author[0000-0001-6395-6702]{Sebastian Gomez}
\affiliation{Space Telescope Science Institute (STScI), 
3700 San Martin Drive, 
Baltimore, MD 21218, USA}

\author[0000-0002-6586-4446]{Alaina Henry}
\affiliation{Space Telescope Science Institute (STScI), 
3700 San Martin Drive, 
Baltimore, MD 21218, USA}
\affiliation{Center for Astrophysical Sciences, Department of Physics and Astronomy, The Johns Hopkins University, 
3400 N Charles St. 
Baltimore, MD 21218, USA}

\author[0000-0003-4512-8705]{Tiger Y.-Y Hsiao}
\affiliation{Center for Astrophysical Sciences, Department of Physics and Astronomy, The Johns Hopkins University, 
3400 N Charles St. 
Baltimore, MD 21218, USA}

\author[0000-0001-6251-4988]{Taylor A. Hutchison}
\affiliation{Department of Physics and Astronomy, Texas A\&M University, College Station, TX, 77843-4242 USA}
\affiliation{George P. and Cynthia Woods Mitchell Institute for Fundamental Physics and Astronomy, Texas A\&M University, College Station, TX, 77843-4242 USA}

\author[0000-0003-4372-2006]{Bethan L. James}
\affiliation{Space Telescope Science Institute (STScI), 
3700 San Martin Drive, 
Baltimore, MD 21218, USA}

\author[0000-0002-8717-127X]{Meridith Joyce}
\affiliation{Space Telescope Science Institute (STScI), 
3700 San Martin Drive, 
Baltimore, MD 21218, USA}

\author[0000-0003-1187-4240]{Intae Jung}
\affiliation{Astrophysics Science Division, NASA Goddard Space Flight Center, Greenbelt, MD 20771, USA}
\affiliation{Department of Physics, The Catholic University of America, Washington, DC 20064, USA}
\affiliation{Center for Research and Exploration in Space Science and Technology, NASA/GSFC, Greenbelt, MD 20771}

\author[0000-0002-3475-7648]{Gourav Khullar}
\affiliation{Department of Astronomy and Astrophysics, University of
Chicago, 5640 South Ellis Avenue, Chicago, IL 60637}
\affiliation{Kavli Institute for Cosmological Physics, University of
Chicago, 5640 South Ellis Avenue, Chicago, IL 60637}

\author[0000-0003-2366-8858]{Rebecca L. Larson}
\altaffiliation{NSF Graduate Fellow}
\affiliation{The University of Texas at Austin, Department of Astronomy, Austin, TX, United States}

\author[0000-0003-3266-2001]{Guillaume Mahler}
\affiliation{Institute for Computational Cosmology, Durham University, South Road, Durham DH1 3LE, UK}
\affiliation{Centre for Extragalactic Astronomy, Durham University, South Road, Durham DH1 3LE, UK}

\author[0000-0001-8057-5880]{Nir Mandelker}
\affiliation{Centre for Astrophysics and Planetary Science, Racah Institute of Physics, The Hebrew University, Jerusalem, 91904, Israel}

\author[0000-0003-0503-4667]{Stephan McCandliss}
\affiliation{Center for Astrophysical Sciences, Department of Physics and Astronomy, The Johns Hopkins University, 
3400 N Charles St. 
Baltimore, MD 21218, USA}

\author[0000-0002-8512-1404]{Takahiro Morishita}
\affiliation{IPAC, California Institute of Technology, MC 314-6, 1200 E. California Boulevard, Pasadena, CA 91125, USA}

\author{Rosa Newshore}
\affiliation{Department of Physics, Clark University Worcester, MA 01610-1477}

\author[0000-0002-5222-5717]{Colin Norman}
\affiliation{Center for Astrophysical Sciences, Department of Physics and Astronomy, The Johns Hopkins University, 
3400 N Charles St. 
Baltimore, MD 21218, USA}
\affiliation{Space Telescope Science Institute (STScI), 
3700 San Martin Drive, 
Baltimore, MD 21218, USA}

\author{Kyle O'Connor}
\affiliation{University of South Carolina, 712 Main St., Columbia, SC 29208, USA}

\author[0000-0001-5851-6649]{Pascal A. Oesch}
\affiliation{Department of Astronomy, University of Geneva, Chemin Pegasi 51, 1290 Versoix, Switzerland}
\affiliation{Cosmic Dawn Center (DAWN), Copenhagen, Denmark}
\affiliation{Niels Bohr Institute, University of Copenhagen, Jagtvej 128, Copenhagen, Denmark}

\author[0000-0003-3484-399X]{Masamune Oguri}
\affiliation{Center for Frontier Science, Chiba University, 1-33 Yayoi-cho, Inage-ku, Chiba 263-8522, Japan}
\affiliation{Department of Physics, Graduate School of Science, Chiba University, 1-33 Yayoi-Cho, Inage-Ku, Chiba 263-8522, Japan}

\author[0000-0002-1049-6658]{Masami Ouchi}
\affiliation{National Astronomical Observatory of Japan, National Institutes of Natural Sciences, 2-21-1 Osawa, Mitaka, Tokyo 181-8588, Japan}
\affiliation{Institute for Cosmic Ray Research, The University of Tokyo, 5-1-5 Kashiwanoha, Kashiwa, Chiba 277-8582, Japan, 3) Kavli Institute for the Physics and Mathematics of the Universe (WPI), University of Tokyo, Kashiwa, Chiba 277-8583, Japan}


\author[0000-0002-9365-7989]{Marc Postman}
\affiliation{Space Telescope Science Institute (STScI), 
3700 San Martin Drive, 
Baltimore, MD 21218, USA}

\author[0000-0002-7627-6551]{Jane R.~Rigby}
\affiliation{Observational Cosmology Lab, NASA Goddard Space Flight Center, Greenbelt, MD 20771, USA}

\author[0000-0003-0894-1588]{Russell E. Ryan Jr}
\affiliation{Space Telescope Science Institute (STScI), 
3700 San Martin Drive, 
Baltimore, MD 21218, USA}

\author[0000-0001-9851-8753]{Soniya Sharma}
\affiliation{Observational Cosmology Lab, NASA Goddard Space Flight Center, Greenbelt, MD 20771, USA}

\author[0000-0002-7559-0864]{Keren Sharon}
\affiliation{Department of Astronomy, University of Michigan, 1085 S. University Ave, Ann Arbor, MI 48109, USA}

\author[0000-0002-6338-7295]{Victoria Strait}
\affiliation{Cosmic Dawn Center (DAWN), Copenhagen, Denmark}
\affiliation{Niels Bohr Institute, University of Copenhagen, Jagtvej 128, Copenhagen, Denmark}

\author[0000-0002-7756-4440]{Louis-Gregory Strolger}
\affiliation{Space Telescope Science Institute (STScI), 
3700 San Martin Drive, 
Baltimore, MD 21218, USA}

\author[0000-0002-0474-159X]{F.X.~Timmes}
\affiliation{School of Earth and Space Exploration, Arizona State University, Tempe, AZ 85287, USA}
\affiliation{Joint Institute for Nuclear Astrophysics - Center for the Evolution of the Elements, USA}

\author[0000-0003-3631-7176]{Sune Toft}
\affiliation{Cosmic Dawn Center (DAWN), Copenhagen, Denmark}
\affiliation{Niels Bohr Institute, University of Copenhagen, Jagtvej 128, Copenhagen, Denmark}

\author[0000-0001-9391-305X]{Michele Trenti}
\affiliation{School of Physics, University of Melbourne, Parkville VIC 3010, Australia}
\affiliation{ARC Centre of Excellence for All-Sky Astrophysics in 3 Dimensions, University of Melbourne, Parkville VIC 3010, Australia}

\author[0000-0002-5057-135X]{Eros Vanzella}
\affiliation{INAF -- OAS, Osservatorio di Astrofisica e Scienza dello Spazio di Bologna, via Gobetti 93/3, I-40129 Bologna, Italy}

\author[0000-0002-4853-1076]{Anton Vikaeus}
\affiliation{Observational Astrophysics, Department of Physics and Astronomy, Uppsala University, Box 516, SE-751 20 Uppsala, Sweden}

\begin{abstract}

The gravitationally lensed star WHL0137-LS, nicknamed Earendel, was identified with a photometric redshift $\zphot = 6.2 \pm 0.1$ based on images taken with the Hubble Space Telescope.
Here we present James Webb Space Telescope (\JWST) 
Near Infrared Camera (NIRCam) images of Earendel
in 8 filters spanning 0.8--5.0$\mu$m.
In these higher resolution images, Earendel remains a single unresolved point source on the lensing critical curve, increasing the lower limit on the lensing magnification to $\mu > 4000$
and restricting the source plane radius further to $r < 0.02$ pc, or $\sim 4000$ AU.
These new observations strengthen the conclusion that Earendel is best explained by an individual star or multiple star system, and support the previous photometric redshift estimate. 
Fitting grids of stellar spectra to our photometry yields a stellar temperature of $T_{\mathrm{eff}} \simeq 13000$--16000 K assuming the light is dominated by a single star.
The delensed bolometric luminosity in this case ranges from $\log(L) = 5.8$--6.6 $L_{\odot}$, which is in the range where one expects luminous blue variable stars.
 Follow-up observations, including \JWST\ NIRSpec scheduled for late 2022, are needed to further unravel the nature of this object, which presents a unique opportunity to study massive stars in the first billion years of the universe.

\end{abstract}


\section{Introduction}

Massive galaxy clusters magnify the distant universe through strong gravitational lensing. 
These cosmic telescopes provide improved spatial resolution over what cutting-edge telescopes can provide alone, allowing the identification of small-scale structures in high redshift galaxies \citep[e.g.,][]{Welch22_clumps,Vanzella22_sunburst,Mestric22}.
In certain cases of precise alignment, galaxy clusters can magnify the light from individual stars by factors of thousands, allowing these stars to be seen above the light of their host galaxies.
The first of these were discovered as transients in images from the Hubble Space Telescope (\HST), at redshifts ranging from $z \sim 1 - 1.5$ \citep[][]{Kelly18,Rodney18,Chen19,Kaurov19_lensedstar}.
Recent discoveries have pushed lensed star observations to greater distances, including recent discoveries at $z = 2.37$ \citep{Diego22_godzilla}, another at $z = 2.65$ with the James Webb Space Telescope (\JWST)  \citep[][]{Chen22_lensedstar}, and a star at $z = 6.2$ discovered in \HST\ imaging \citep[][]{Welch2022_earendel}.

\JWST\ \citep{Gardner06_JWST}, which has recently completed commissioning and begun science operations \citep{Rigby22_JWSTcommissioning}, will continue improving our ability to study distant lensed stars in detail. 
Besides already discovering new lensed stars \citep[][]{Chen22_lensedstar}, \JWST\ will enable more detailed study of the highest redshift lensed stars. 
The combination of this powerful new observatory and gravitational lensing could also be our best chance at observing Population III stars directly \citep{Windhorst18}.

In this paper, we present new \JWST\ imaging of the lensed star Earendel \citep[RA = 01:37:23.25, Dec = $-$8:27:52.27;][]{Welch2022_earendel}.
Earendel is in a $\zphot = 6.2 \pm 0.1$ galaxy dubbed the Sunrise Arc,
the most highly magnified galaxy at $z \sim 6$ \citep{Salmon2020}.
It is lensed by a massive  $z = 0.566$ galaxy cluster WHL J013719.8--082841 (henceforth \WHL; RA = 01:37:25.0, Dec = $-$08:27:23, J2000;
\citealt{WHL12,WenHan15}).
We describe the \JWST\ imaging data in Section \ref{sec:data}.
The photometric redshift estimates for the Sunrise Arc are presented in Section \ref{sec:photoz}, and the updated magnification and size constraints are described in Section \ref{sec:magnif}.
Our photometric temperature estimate is discussed in Section \ref{sec:temp}.
We investigate possible variability of the source in Section \ref{sec:variable}.
Our results are presented and contextualized in Section \ref{sec:results}.
Finally, we end with our conclusions in Section \ref{sec:conclusions}.
Throughout, we assume a flat $\Lambda$CDM cosmological model, with $\Omega_{\rm M} = 0.3$, $\Omega_{\Lambda} = 0.7$, and the Hubble constant $H_0 = 70~ \text{km s}^{-1}\text{Mpc}^{-1}$.

Data products, reduced images, lens models, and anlaysis code are available via our website.\footnote{\url{https://cosmic-spring.github.io}}

\begin{figure*}
    \centering
    \includegraphics[width=0.9\textwidth]{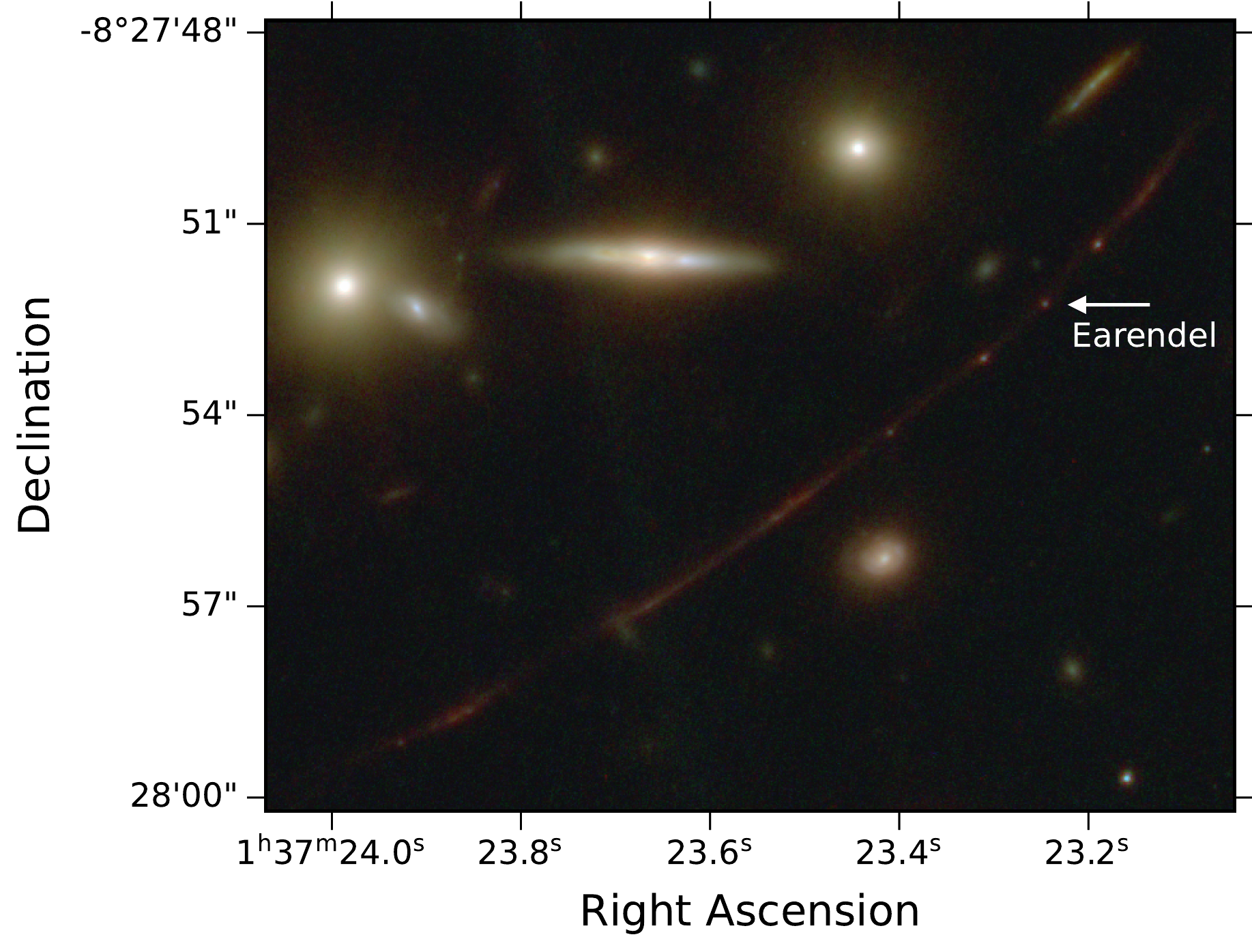}
    \caption{\JWST\ NIRCam image of the $z \sim 6.2$ Sunrise Arc, including the lensed star Earendel marked with an arrow. 
    This $15\farcs2 \times 12\farcs4$ color image combines all 8 NIRCam images on a $0\farcs02$ pixel scale.}
    \label{fig:color}
\end{figure*}

\section{Data}
\label{sec:data}

\begin{table}[]
    \centering
    \begin{tabular}{c c c c c}
        Filter & $\lambda$ & Exp. & Flux Density & \\
         & ($\mu$m) & Time (s) &   (nJy) & AB mag \\
         \hline
         F090W & 0.8--1.0  & ~\,2104  & $32 \pm 5$ & $27.77 \pm 0.19$   \\
         F115W & 1.0--1.3  & ~\,2104  & $57 \pm 7$ & $27.05 \pm 0.14 $  \\
         F150W & 1.3--1.7  & ~\,2104  & $50 \pm 6$ & $27.20 \pm 0.14$  \\
         F200W & 1.7--2.2  & ~\,2104  & $43 \pm 3$ & $27.46 \pm 0.09$  \\
         F277W & 2.4--3.1  & ~\,2104  & $63 \pm 5$ & $26.90 \pm 0.08$  \\ 
         F356W & 3.1--4.0  & ~\,2104 & $66 \pm 4$ &  $26.86 \pm 0.06$ \\
         F410M & 3.8--4.3  & ~\,2104 & $64 \pm 5$ &  $26.89 \pm 0.08$ \\
         F444W & 3.8--5.0  & ~\,2104 & $62 \pm 6$ &  $26.83 \pm 0.11$ 
    \end{tabular}
    \caption{\JWST\ NIRCam photometry of Earendel in 8 filters, measured as described in Appendix \ref{phot_appendix}.}
    \label{tab:photometry}
\end{table}

\subsection{\JWST\ NIRCam}

\begin{figure*}
    \centering
    \includegraphics[width=0.9\textwidth]{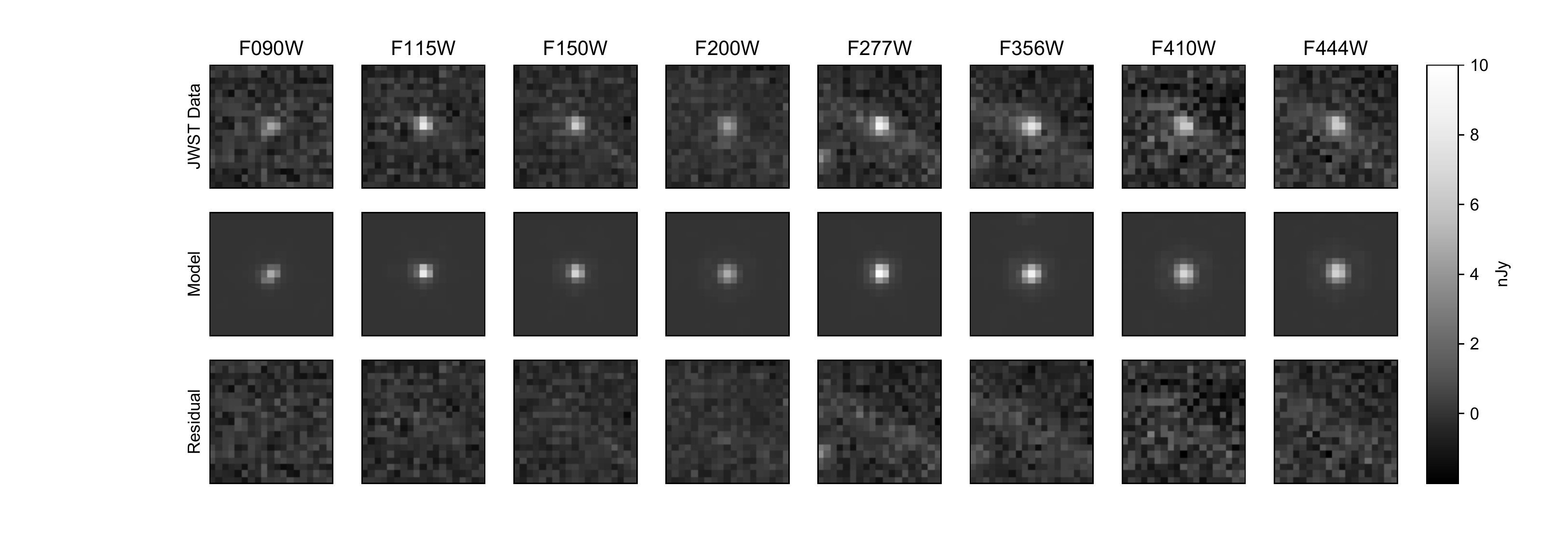}
    \caption{ Image cutouts of Earendel in each \JWST\ filter are shown in the top row in native pixels (SW 0\farcs031; LW 0\farcs063). Individual exposure cutouts are fit with a PSF point source model, and the sum of these 4 exposure-level models are shown in the middle row. The residuals are also calculated on the level of individual exposures, then summed to produce the bottom row of this figure. The residuals are consistent with noise in each filter, indicating that Earendel is a point-like source and supporting the interpretation that it is a distant lensed star.
    }
    \label{fig:stamps}
\end{figure*}

Earendel was first identified in \HST\ imaging taken as part of the Reionization Lensing Cluster Survey \citep[RELICS; GO 14096][]{Coe19_relics} and a follow-up program (GO 15842, PI Coe), as described in \cite{Welch2022_earendel}.
We recently obtained additional imaging from the newly commissioned \JWST\ NIRCam instrument as part of Cycle 1 GO program 2282 (PI Coe).
These images span a wavelength range of 0.8--5 $\mu$m in eight filters, presented in Table \ref{tab:photometry}. 
A color image of the Sunrise Arc hosting Earendel is shown in Figure \ref{fig:color}, and image stamps of Earendel in each filter are shown in Figure \ref{fig:stamps}.
Each filter was observed for a total of 2104 seconds of exposure time.
We utilized four dithers to cover the 5\arcsec\ gaps between the short wavelength (SW; $\lambda < 2.4 \mu$m) detectors, as well as improving the resolution of our final drizzled images and minimizing the impact of image artifacts and bad pixels. 
Additional imaging in four filters (F090W, F115W, F277W, F356W) and NIRSpec spectroscopy for GO 2282 is expected in December 2022.

We retrieved \JWST\ pipeline\footnote{\url{https://jwst-pipeline.readthedocs.io}}
Level 2b data products ({\tt cal.fits}) from 
MAST\footnote{\url{https://mast.stsci.edu}}.
They included updated zeropoints based on in-flight data,
delivered to CRDS version 11.16.3 
with {\tt jwst\_0942.pmap} reference files 
on 2022-07-29, the day before these observations.

We processed the \JWST\ Level 2 data products using the \texttt{grizli} pipeline\footnote{\url{https://github.com/gbrammer/grizli}} \citep{Brammer21_grizli}.
This processing pipeline reduces striping from 1/f noise and masks ``snowballs'' in the images, and includes a zeropoint correction based on observations of the LMC\footnote{\url{https://github.com/gbrammer/grizli/pull/107}}. These corrections match the later {\tt jwst\_0989.pmap} reference file zeropoints to within a few percent. 
All images are then aligned to a common WCS registered to GAIA DR3 \citep{Gaia_EDR3}.
The pipeline next drizzles the images to a common pixel grid of 0\farcs04 per pixel using the \texttt{astrodrizzle} software \citep{MultiDrizzle,AstroDrizzle,DrizzlePac}.
The short wavelength NIRCam images are drizzled to a higher resolution grid of 0\farcs02 per pixel, aligned to the lower resolution grid (with each low-resolution pixel corresponding to $2\times2$ high-resolution pixels). 
These reprocessed images are publicly available via our website.


Sources are then detected in a weighted sum of the drizzled NIRCam images in all filters using a Python implementation of SourceExtractor called SEP \citep{Barbary2016,sextractor}. 
Fluxes are then calculated for each source in three circular apertures, 0\farcs36, 0\farcs5, and 0\farcs7.
The 0\farcs5 aperture fluxes are used for the photo-$z$ calculations with \texttt{EAZY}\footnote{\url{https://github.com/gbrammer/eazy-photoz}} \citep{Brammer08_eazy}, described below.

The source extraction parameters utilized in the \texttt{grizli} pipeline blend multiple features of the Sunrise Arc together.
In order to extract reliable, uncontaminated fluxes for Earendel, we therefore perform additional photometric measurements on the object directly, using a variety of measurements.
Distinguishing between flux originating from Earendel and flux originating from the host arc is non-trivial.
While photometry of isolated point sources is traditionally well measured with packages such as \texttt{photutils} \citep{photutils} or Source Extractor \citep{sextractor}, this case is somewhat more complex given the point source is embedded within the Sunrise Arc. 
Slight differences in how backgrounds are subtracted and how the PSF is determined can produce non-negligible differences in the resulting fluxes. 
We thus choose to measure the fluxes from a variety of methods, and produce final values from the average of each method.
We utilize both aperture photometry and PSF-matched photometry, with various aperture sizes and PSF models, measured by ten independent observers as described in Appendix \ref{phot_appendix}.
The resultant fluxes are presented in Table \ref{tab:photometry}.
This consensus photometry approach allow us to understand the range of possible fluxes given different assumptions, thus incorporating systematic uncertainties into our final measurement.

\subsection{\HST\ WFC3/IR}

We additionally study \HST\ data taken as part of both the original follow-up program (GO 15842) and an ongoing monitoring program (GO 16668, PI Coe). 
The goal of the monitoring program is to assess variability in the lensed star, thus it observes in the WFC3/IR F110W bandpass, matching the strongest detection of the GO 15842 program.
In total this monitoring program will obtain four additional epochs, however only two have been observed thus far.
The observations of GO 15842 occurred on November 4, 2019, and November 27, 2019.
The first two epochs of GO 16668 observations occurred on November 28, 2021, and January 29, 2022.
These observations currently span just over two years.

\section{Photometric Redshift Estimate} \label{sec:photoz}

We perform initial photometric redshift estimation using \texttt{EAZY}, utilizing a set of galaxy spectral templates generated with FSPS \citep{Conroy09,Conroy10,ConroyGunn10}. 
Redshifts are allowed to vary over the range $0.01 \leq \zphot \leq 18$, in steps of 0.01. 
A redshift prior is applied based on previously observed galaxy count rates as a function of redshift and magnitude in the \HST\ F160W filter.

The \texttt{EAZY} calculations are performed as part of the image reduction pipeline, and thus are run on each segment identified by Source Extractor within said pipeline. 
This splits the Sunrise Arc up into a total of 4 components, and gives photometric redshifts of $\zphot = 6.00^{+0.09}_{-0.11}$, $\zphot = 6.31^{+0.08}_{-0.15}$, $\zphot = 6.04^{+0.20}_{-0.23}$, and $\zphot = 6.25^{+0.12}_{-0.16}$ for each component. 
Combining these redshifts yields a total redshift estimate for the full arc of $\zphot = 6.15^{+0.27}_{-0.34}$. 

Previous works \citep{Salmon2020,Welch2022_earendel} have used other SED fitting methods to estimate the photometric redshift of the Sunrise Arc using only the \HST\ imaging.
These results consistently found redshifts $\zphot = 6.2 \pm 0.1$, consistent with the present fit. 
We thus adopt a fiducial redshift of $z = 6.2$ for the Sunrise Arc, and by extension Earendel, which we use for all further calculations in this paper.

\section{Magnification and Size Constraints} \label{sec:magnif}

For the present analysis, we utilize the five previously published lens models presented in \cite{Welch2022_earendel}. 
These models were made using four lens modeling software packages, Light-Traces-Mass \citep[LTM,][]{Zitrin09,Zitrin15,Broadhurst05}, Glafic \citep{Oguri2010}, WSLAP+ \citep{Diego05wslap,Diego07wslap2}, and Lenstool \citep{JulloLenstool07,JulloLenstool09}.
A factor of 6 variation exists between the slope of the lensing potential in these models, adding considerable uncertainty to our magnification and maximum size constraints. 

We constrain the magnification of the lensed star following a similar procedure to that described in \cite{Welch2022_earendel}.
We first observe that Earendel is consistent with being a point source in each of the \JWST\ filters, as shown in Figure \ref{fig:stamps}.
We model the object as a point source using the four individual exposures (\texttt{cal} files) for each filter.
The point source is convolved with an empirically derived PSF model measured from individual \texttt{cal} exposures of the Large Magellanic Cloud (LMC) calibration field (Anderson et al. 2022, in prep).
These exposure-level PSF models appear to be more accurate than empirical PSF models based on the final drizzled images.
The point source model is then subtracted from the individual exposure, creating a total of four residual images.
These residuals are then summed, centering on Earendel's centermost pixel in each exposure, to create the full residuals presented in Figure \ref{fig:stamps}.
The residuals are consistent with noise in each filter, indicating that Earendel is indeed a point-like source.

We constrain the magnification following the method of \cite{Welch2022_earendel}. 
Briefly, we measure the maximum separation of two point sources that would remain unresolved, then calculate the minimum magnification using the relation $\mu = \mu_0 / D$, where $D$ is the distance to the critical curve in arcseconds, and $\mu_0$ is a constant that can be fit for each lens model \citep{Diego19}.
We find that the distance between two resolved simulated point sources ($2\xi$ in their notation) is about one native pixel, which in the case of \JWST\ is 0.031\arcsec.
We then use this magnification, along with the measured image plane size, to calculate the maximum possible size of the object in the source plane, again following the method of \cite{Welch2022_earendel}.


Using this technique, we are able to improve our constraints on the magnification and maximum radius of Earendel. 
We find best-fit magnification values ranging from 6000--35000 depending on the lens model, while the lower limit including uncertainties is $\mu \geq 4000$ (see Table \ref{tab:mag_radius}).
The updated magnifications along with the higher spatial resolution images allow us to set tighter constraints on the radius as well, with maximum radii ranging from $r < 0.005$-0.02 pc (1000-4000 AU), depending on the lens model.

\section{Temperature Estimate} \label{sec:temp}

\begin{figure*}
    \centering
    \includegraphics[width=0.45\textwidth]{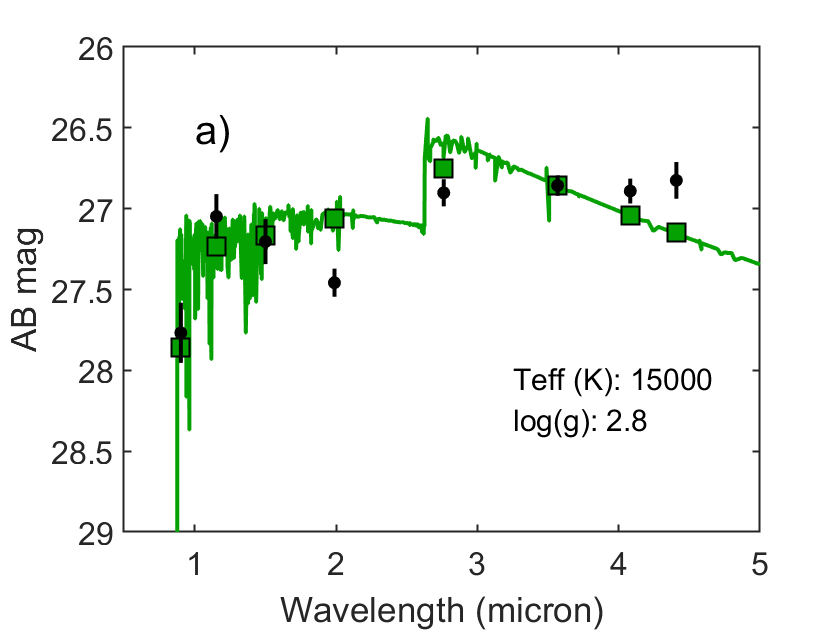}
		\includegraphics[width=0.45\textwidth]{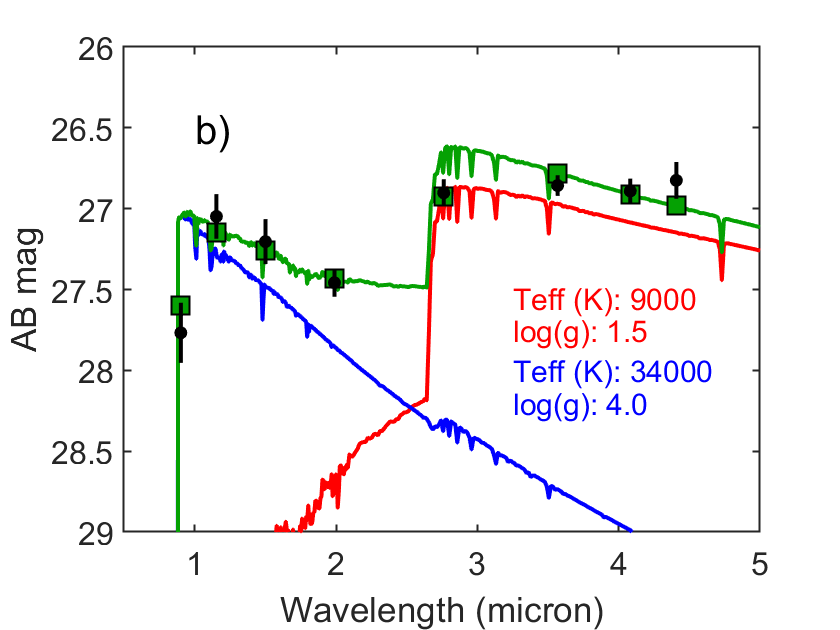}
    \caption{SED fits to the observed JWST/NIRCam photometry of Earendel (black circles). {\bf a)} Single-star fit using a $T_\mathrm{eff}=15000$ K, SMC metallicity ($\sim 1/7$ solar), $\log(g)=2.8$ stellar atmosphere spectrum from the PoWR OB-star grid \citep{Hainich19} with moderate mass loss ($10^{-7.58}\ M_\odot yr^{-1}$ for this particular model)\ redshifted to $z=6.2$ (green spectrum, with green squares indicating the filter-integrated fluxes). Under the assumpton of $2\mu=35000$ predicted by the LTM model, this star would need to have $\log (L_\mathrm{bol}/L_\odot)\approx 5.8$ to match the observed SED. 
		{\bf b)} Double-star fit to Earendel's JWST/NIRCam photometry using two stars from the \citet{Lejeune97} stellar atmosphere grid redshifted to $z=6.2$. Here, the cooler star (red spectrum) has $T_\mathrm{eff}=9000$ K, $\log(g)=1.5$ and $[\mathrm{M/H}]=-1$, whereas the hotter star (blue spectrum) has $T_\mathrm{eff}=34000$ K, $\log(g)=4.0$ and $[\mathrm{M/H}]=-1$. For this particular combination of stars, and under the assumption that $2\mu=35000$ (LTM) holds for both stars, the colder star would have $\log (L_\mathrm{bol}/L_\odot)\approx 5.3$ and the hotter $\log (L_\mathrm{bol}/L_\odot)\approx 5.8$. The green spectrum indicates the sum of the flux from the two stars and the green squares the corresponding filter-integrated fluxes.}\label{fig:stellarSED}
\end{figure*}

The JWST/NIRCam photometry of Earendel, presented in Table~\ref{tab:photometry}, is not easily fitted by local low-mass stars, brown dwarfs or giant exoplanets in the Milky Way, based on the theoretical spectral energy distributions (SEDs) of \citet{Baraffe15} and \citet{Phillips20} for such objects. If we instead assume that the light from Earendel comes from a single, highly magnified star at high redshift, then a broad scan of SED fits based on the stellar atmosphere set of \citet{Lejeune97}\footnote{The part of the \citet{Lejeune97} compilation most relevant for the Earendel analysis is that based on Kurucz ATLAS stellar atmosphere models, modified to fit empirical color-tempreature calibrations.}, assuming no dust reddening and zero transmission of flux through the IGM shortward of the redshifted Lyman-$\alpha$ line, favour $z\approx  5.7$--6.5 and B-type stars with $T_\mathrm{eff}\approx 13000$--16000 K. Metallicity only has a minor impact on the quality of the fit and cannot be constrained by the data.

Assuming $z=6.2$ and refining the fit using the more realistic B-star TLUSTY stellar atmosphere grid by \citet{Lanz07} or the Potsdam Wolf-Rayet (PoWR) model atmosphere grid for OB stars with various levels of mass loss \citep{Hainich19} results in best-fitting temperatures of $T_\mathrm{eff}= 15000$ K (at the lower $T_\mathrm{eff}$ limit of either grid). 

In Fig.~\ref{fig:stellarSED}a, we present the best fit to the SED of Earendel allowed by the PoWR stellar atmosphere model at SMC metallicity \citep[$\sim 1/7$ solar, broadly consistent with the fiducial metallicity adopted by][]{Welch2022_earendel}. With both this model and similar fits produced using TLUSTY or the \citet{Lejeune97} set, the shift in flux between F200W and the longer-wavelength filters is interpreted as due to the Balmer break. However, even the best-fitting models are unable to reproduce the size of this break (as evident from the significant offset between model and observational data in F200W) or the observed flux in the longest-wavelength filters, and the resulting $\chi^2$ is very high ($\chi^2 \approx 38$ for the plotted fit).  

The primary reason why these SED models struggle to provide a convincing fit to the photometric data of Earendel is that the observed SED appears to feature both a relatively strong Balmer break (typical of $T_\mathrm{eff}\lesssim 13000 K$ stars) and a steep ultraviolet continuum slope (typical of $T_\mathrm{eff}\gtrsim 20000 K$ stars). Adding the effects of dust at the redshift of Earendel would not significantly improve the fit, since dust reddening would preferentially affect the ultraviolet continuum and require an intrinsically bluer (hotter) star to match the observed data, thereby increasing the tension with the Balmer break strength. Since surface gravity ($\log(g)$) modifies the slope of the ultraviolet continuum at fixed $T_\mathrm{eff}$, the SED of Earendel makes the $\chi^2$-minimization procedure favour higher $\log(g)$ than would be expected for a single very massive star. For example, the PoWR fit presented in Fig.~\ref{fig:stellarSED}a has $\log(g)=2.8$, which would correspond to  stars with ZAMS mass $\approx 10$\ $M_\odot$ based on the stellar evolutionary models of \citet{Szecsi22}. Assuming the LTM magnification of $2\mu=35000$, the model presented in Fig. ~\ref{fig:stellarSED}a corresponds to a bolometric luminosity of $\log(L/L_\odot)\approx 5.8$, which at this temperature would be more typical of an evolved star with ZAMS mass $\approx 40\ M_\odot$. Rejecting high-$\log(g)$ models only acts to further degrade the fit, since such models exhibit less steep ultraviolet continuum slopes. Lens models with lower magnifications would not solve the problem either, since these require even higher bolometric luminosities, and thus higher ZAMS masses.
However, the $\log(g)$ tension can be reduced somewhat by assuming that  Earendel is composed of several stars with approximately the same temperature, in which case the inferred luminsity could be explained by $\approx 2$ stars of ZAMS mass $30\ M_\odot$, or $\approx 4$ stars of ZAMS mass $20\ M_\odot$.

While photometric uncertainties remain large, it is tempting to consider the possibility at least two stars of different temperatures are contributing significantly to the observed SED of Earendel, since this could potentially explain the presence of both a steep ultraviolet continuum slope and a strong Balmer break. In Fig.~\ref{fig:stellarSED}b, we provide an example of such a double-star fit, in which the summed contributions from one star with $T_\mathrm{eff}=9000$ K star and one with $T_\mathrm{eff}=34000$ K provide a good fit ($\chi^2  \approx 5$) to the observed SED of Earendel. However, given the limited number of photometric data points available and the number of free parameters involved in such double-star fits, the parameters of the two stars are not well constrained. If two stars are involved, their magnifications may also differ, which further complicates the analysis. We therefore defer a detailed analysis of such scenarios until spectroscopic data is available.

Another possible explanation for the somewhat puzzling SED of Earendel is that some of the observed filter fluxes are affected by emission lines (e.g. CIV 1549 in F115W, [OIII]5007 and H$\beta$ in F277W, H$\alpha$ in F444W) from either a wind surrounding this object or from a more extended and less magnified HII region produced either by Earendel itself or by other massive stars in its surrounding. Upcoming JWST/NIRSpec spectroscopy of Earendel should make it possible either detect or set strong upper limits on the contribution of such lines.

\section{Variability} \label{sec:variable}

\begin{figure}
    \centering
    \includegraphics[width=0.45\textwidth]{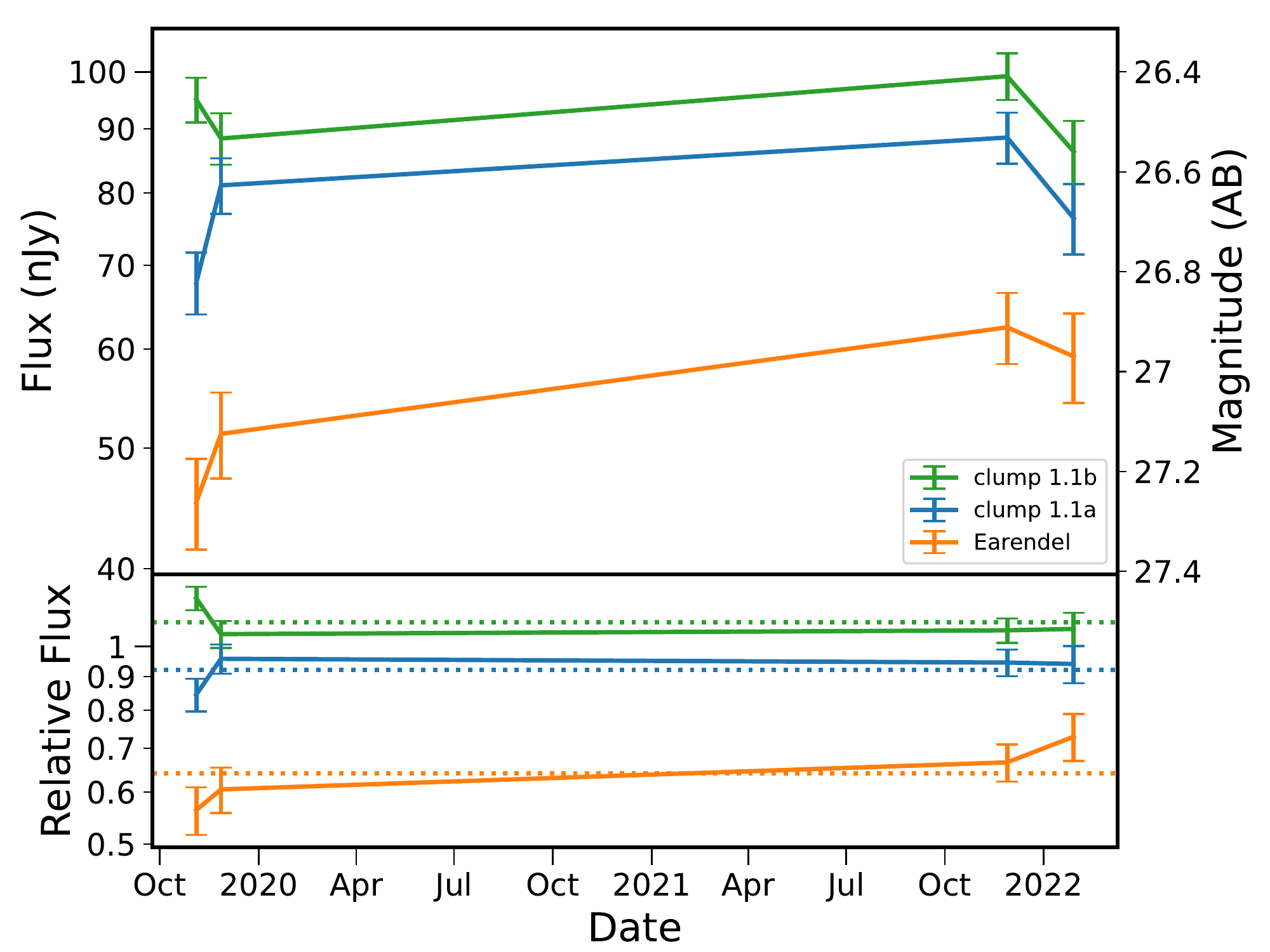}
    \caption{\HST\ WFC3/IR F110W fluxes measured over four epochs, spanning over two years,
    for Earendel and two nearby lensed mirror images of a star cluster in the Sunrise Arc.
    The bottom panel shows fluxes normalized to the geometric mean of the star clusters,
    which we assume remain at constant flux 
    (not variable and unaffected by microlensing).
    No significant time variation is observed within the uncertainties.
    }
    \label{fig:variable}
\end{figure}

Gravitationally lensed stars typically experience fluctuations in their overall magnifications due to microlensing \citep[e.g.,][]{Kelly18,Chen19,Rodney18,Chen22_lensedstar,Diego22_godzilla}.
Earendel was found to be somewhat unique, as its microlensing configuration lends itself to modest variation over time \citep{Welch2022_earendel}.
To further monitor the variability of Earendel's magnification, we have been repeating observations using \HST\ in the same filter (WFC3 F110W). 
These repeat observations offer the best chance to look for variability, as using different filters introduces additional uncertainty that can obscure true changes. 

To ensure consistency between flux measurements in each epoch, we measure the brightness within a common circular aperture with radius 0\farcs3 = 5 pixels in each drizzled \HST\ image.
We utilize a common circular annulus immediately outside the central aperture to measure a local background, which is then subtracted from the central aperture flux. 
The resulting flux measurements are shown in Figure \ref{fig:variable}, and the flux values are given in Table \ref{tab:variable}. 
As a comparison, we repeat this flux measurement for each of the mirror-imaged clumps that bracket Earendel \citep[1.1a/1.1b in the notation of][]{Welch2022_earendel}. 
These are also plotted and tabulated in Fig. \ref{fig:variable} and Table \ref{tab:variable}.

\begin{table}[]
    \centering
    \begin{tabular}{c c c c}
         & Earendel & clump 1.1a & clump 1.1b \\
        Obs. Date & Flux (nJy) & Flux (nJy) & Flux (nJy) \\
        \hline 
        Nov.~4 2019  & $45 \pm 4$ & $68 \pm 4$ & $95 \pm 4$ \\
        Nov.~27 2019 & $51 \pm 4$ & $81 \pm 4$ & $88 \pm 4$\\
        Nov.~28 2021 & $62 \pm 4$ & $89 \pm 4$ & $99 \pm 4$\\
        Jan.~29 2022 & $59 \pm 5$ & $76 \pm 5$ & $86 \pm 5$
    \end{tabular}
    \caption{\HST\ F110W flux values measured in four epochs over a two year period for Earendel and mirror images of a nearby lensed star cluster, as plotted in Figure \ref{fig:variable}.}
    \label{tab:variable}
\end{table}

We find that the highest and lowest flux values for Earendel differ by $\sim 3 \sigma$. 
We also find that the largest deviation from the median of all measurements is $2.7 \sigma$. 
While this is enough of a difference to warrant further investigation, we cannot conclude that these values differ in a statistically significant way. 
Additional epochs and deeper imaging would better constrain the time variability of this object.

The current lack of clear variability is consistent with the microlensing analysis of \citet{Welch2022_earendel}, which predicts that the magnification should generally stay consistent within a factor of 2.
Our present variation is at most a factor of $\sim 1.4$. 
The lack of clear variation may also be indicative of more than one star being present. 
Each star would then cross microcaustics at different times, minimizing the effect on the total flux of such microcaustic crossing events.
Ultimately, additional epochs of observation with greater signal to noise will be required to fully understand the variability of Earendel.

\section{Discussion}
\label{sec:results}

\begin{figure*}
    \centering
    \includegraphics[width=0.9\textwidth]{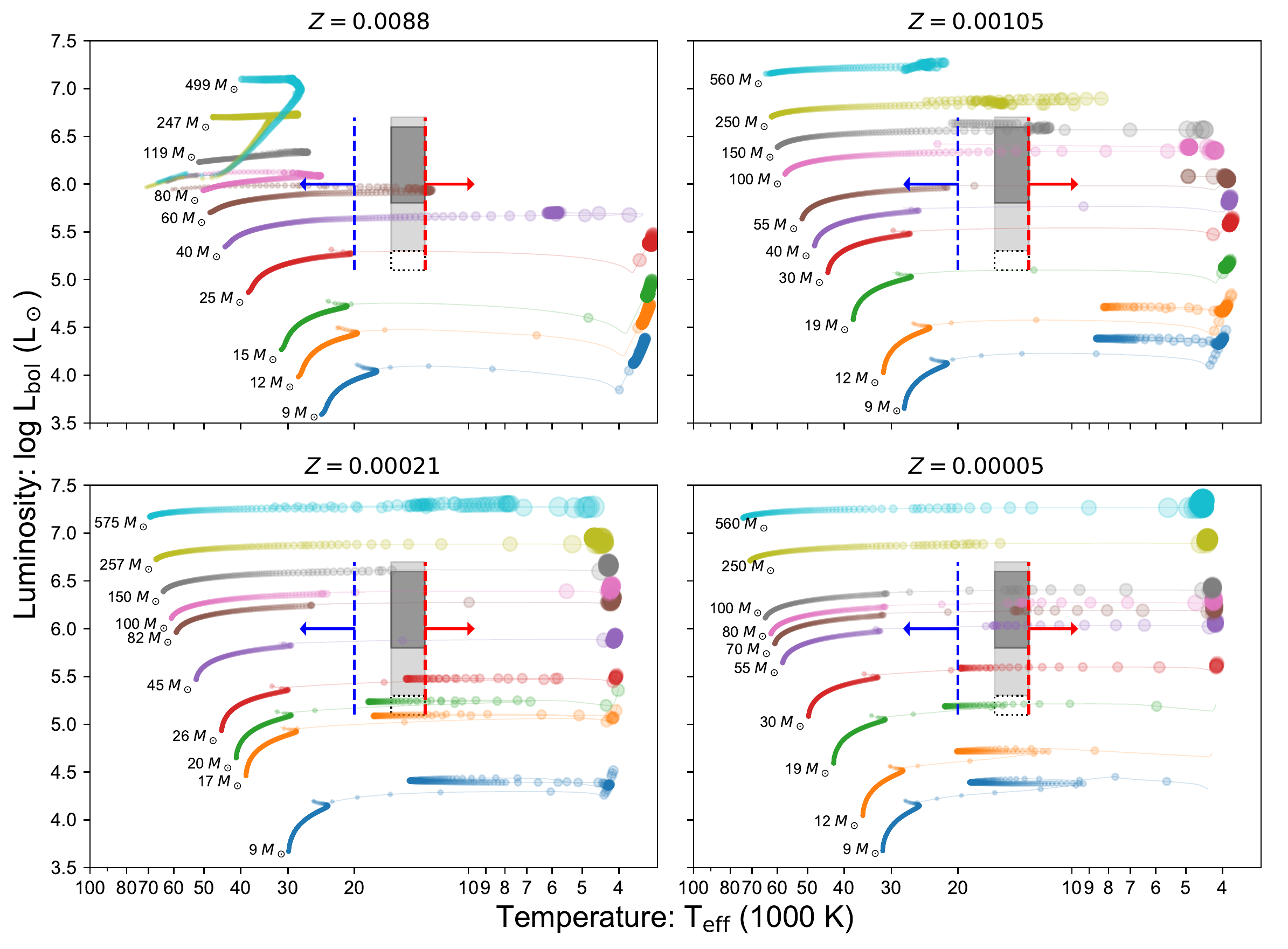}
    \caption{HR Diagram showing best-fit temperature and luminosity constraints for a single star in grey alongside model stellar evolution tracks by \cite{Szecsi22} which account for stellar wind mass loss and rotation, see their paper for details. We show metallicities ranging from $Z = 0.0088$ (top left panel, about $0.7 Z_{\odot}$) to $Z = 0.00005$ (bottom right panel, about $Z_{\rm SMC}/50$).
   The darker grey region covers the range of best-fit magnifications from our five lens models, while the fainter grey region includes the uncertainties on each lens model. The low-luminosity region indicated by the dotted box is allowed within the lens model uncertainties, but exceeds the $\mu = 10^5$ magnification upper limit inferred from microlensing measurements in \cite{Welch2022_earendel}. 
   The red and blue dashed lines indicate rough limits for a two-star solution, which would require a hotter star $> 20000$ K to fit the blue UV slope, and a cooler star $< 13000$ K to fit the observed Balmer break. The large parameter space of two star solutions coupled with our few photometric data points means we cannot fully constrain these solutions, hence they are shown as rough limits only.
       }
    \label{fig:HRD}
\end{figure*}

\begin{table*}[]
    \centering
    \begin{tabular}{c c c c c c c}
        Lens Model & $\mu_0$ & $D_{\text{crit}}$ &  Magnification & Radius & Luminosity & $M_V$ \\
         & & (\arcsec) & $2\mu (\times 10^3)$ & (pc/AU) & ($\log(L_{\odot})$) & (AB, rest frame) \\
        \hline 
        LTM          & 113 & 0.006 & $35^{+140}_{-10}$    & $<0.005$ (1020 AU) & $5.8^{+0.1}_{-0.7}$ & $-8.5^{+1.6}_{-0.3}$ \\
        Glafic (c=1) & 69  & 0.005 & $28^{+113}_{-8}$ & $<0.008$ (1600 AU) & $5.9^{+0.1}_{-0.7}$ & $-8.8^{+1.7}_{-0.3}$ \\
        Glafic (c=7) & 23  & 0.005 & $9^{+38}_{-3}$   & $<0.012$ (2520 AU) & $6.4^{+0.1}_{-0.7}$ & $-10.0^{+1.7}_{-0.3}$ \\
        WSLAP        & 28  & 0.009 & $6^{+26}_{-2}$   & $<0.019$ (3890 AU) & $6.5^{+0.2}_{-0.7}$ & $-10.4^{+1.7}_{-0.3}$ \\
        Lenstool     & 18  & 0.006 & $6^{+23}_{-2}$   & $<0.020$ (4160 AU) & $6.6^{+0.1}_{-0.7}$ & $-10.5^{+1.7}_{-0.3}$
    \end{tabular}
    \caption{Magnification and delensed parameter measurements for each lens model. Luminosities and V-band absolute magnitudes are calculated assuming a single star dominates the observed flux.}
    \label{tab:mag_radius}
\end{table*}

The \JWST\ imaging observations presented herein support the conclusion of \cite{Welch2022_earendel} that the object nicknamed Earendel is an extremely magnified star at redshift $\zphot = 6.2$. 
The increased spatial resolution of \JWST\ allows us to improve our constraints on the total magnification of the star, resulting in best-fit values ranging from 6000 to 35000 depending on the lens model (see Table \ref{tab:mag_radius}). 
The lower limit on the magnification including all uncertainties has increased by a factor of 4, from 1000 to 4000, thanks to improved constraints on the distance to the critical curve. 

The increased spatial resolution also improves the constraints on the maximum size of Earendel in the source plane. 
Whereas \cite{Welch2022_earendel} found upper limits on source plane radius ranging from 0.09--0.36 pc, we now find maximum radii $r < 0.005$--0.02 pc (1000--4000 AU), depending on the lens model.
This further distinguishes Earendel from known young massive star clusters, which have typical radii of $\sim 1$ pc \citep{Portegies-zwart2010_YMCs}.
Even small central cores observed in nearby star clusters such as R136 do not reach below tenths of a parsec \citep[e.g.,][]{crowther16_r136}, which is still far larger than our measured radii.
This strengthens our conclusion that Earendel is most likely an individual star system. 

While our new radius constraints conclusively rule out a star cluster, the tightest constraint of $<1000$ AU leaves room for multiple companion stars.
Massive stars in the local universe often have companions, and frequently have more than one companion.
Secondary companions are typically located at a median distance of less than 2 AU, while tertiary companions are found at $\sim20$ AU (\citealt{sana12,Sana14}, and Fig~3 in the review by \citealt{offner22}). 
A multiple star system would thus remain unresolved in our imaging.

For our analysis of the stellar properties of Earendel, we first assume that the light is either coming from a single star, or it is dominated by the brightest star of a compact group of stars.
This gives a best-fit temperature range of $T_\mathrm{eff} = 13000$--16000 K. 
We calculate intrinsic bolometric luminosities based on the magnifications given by our various lens models, finding a best-fit range of $\log(L) = 5.8$--$6.6 L_{\odot}$ (see Table \ref{tab:mag_radius}).
We overplot these temperature and luminosity constraints on the Hertzsprung-Russel (HR) diagram in Figure \ref{fig:HRD}, alongside stellar evolution models of varying metallicities from \cite{Szecsi22}. The dots represent evenly spaced timesteps of 10,000 years, giving an indication of how fast the star is evolving through the diagram. These models account for stellar wind mass loss scaled down with metallicity following \citet{vink01} and assume that the stars are born with a modest rotation of 100 km\,s$^{-1}$. Models with much faster rotation (not shown here) typically stay too hot to explain the temperature derived for Earendel. 
 
Taking our range of magnification estimates at face value, this gives us a range of single star ZAMS masses of 20--200 $M_{\odot}$. The higher metallicity models prefer solutions where the star is at the end of its main sequence stage. The lowest metallicity models also allow for central helium-burning solutions. 
We note that the high ZAMS mass and high luminosity range is in line with observational biases expected for lensed stars, which favor observations of more luminous O- and B-type stars over fainter ones \citep{Meena22}.

There are several important caveats that effectively shrink this range. First, the microlensing analysis presented in \cite{Welch2022_earendel} indicates that the highest achievable magnification is likely around $\mu = 100000$, while our smooth lens models alone allow for magnifications up to $\mu = 175000$. 
This makes the lowest-mass range somewhat dubious, as microlensing could limit the magnification enough to make observations of stars this faint unlikely.
Furthermore, the probability of an object achieving a given magnification falls proportional to the magnification squared, $P(>\mu) \propto \mu^{-2}$.
Our high-end magnification estimate of 35000 is therefore 25 times less likely to occur than a magnification of 7000.
While this does not rule out such high magnifications, it favors more luminous stars at lower magnifications.

The high mass, high luminosity end comes with a caveat as well. Theoretical models of the evolution of such high-mass stars are still plagued by substantial uncertainties, which particularly affect how long a star may spend in which part of the HR diagram. The models shown here account for substantial mixing beyond the convective core \citep[as detailed in][]{brott11}, but even larger amounts of mixing may be needed \citep[e.g.\ ][]{vink01}, which would extend the end of the main sequence to cooler temperatures. 
At a temperature of 15000 K, luminosities above $\sim 10^6 L_{\odot}$ would exceed the Humphreys-Davidson limit, an empirical limit above which almost no stars are found to exist at least for stellar populations in the local universe \citep{HumphreysDavidson78_I,HumphreysDavidson79_III}.
Stars living in this regime tend to be Luminous Blue Variables \citep[LBVs, e.g.\ ][]{Smith04}, bright stars that experience irregular eruptive episodes of mass loss. 
The physical mechanism for these eruptions is not fully understood \citep[see, however,][]{Jiang18}. 
In particular, the question of whether the LBV phenomena still occurs at low metallicity is a matter of debate \citep[e.g.\ ][]{Davies18, Kalari18}.
If indeed the star has been experiencing strong mass loss, one may expect a dense outflow. The photosphere may then not be located at the hydrostatic layer but further out in the dense stellar wind. This may mean that the actual star is hotter than the temperature we have inferred here.  
We note that the \cite{Szecsi22} stellar models do not include episodic eruptions of mass as observed for LBVs, so these stellar tracks may overpopulate the region in the HR diagram above the Humphreys Davidson limit. 

A further interesting question related to the LBV possibility is whether Earendel shows signs of  variability, for example from possible eruptions. 
Current data hints at possible variations, but no statistically significant variability has yet been observed. 
Follow-up observations are ongoing to further explore this possibility.

If instead Earendel is made up of a tightly bound group of stars with similar temperatures, the intersection with the Humphreys-Davidson limit could be avoided. 
For example, as mentioned in Section \ref{sec:temp}, two stars with ZAMS mass $\sim 30 M_{\odot}$ and roughly equal temperatures could produce a similar result to our single star SED fit.
However, there are several discrepancies between our best-fit single temperature model and the measured photometry. 
In particular, the model spectra used do not fully reproduce the F200W--F277W color (0.56 mag), which we interpret as the 4000\AA ~Balmer break.
The cooler ($T_{\mathrm{eff}} \lesssim 13000$ K) stars that best fit the Balmer break then struggle to reproduce the apparently blue UV slope, which would favor stars with $T_{\mathrm{eff}} \gtrsim 20000$ K. 
We therefore consider possible two star solutions, and present one such solution in Figure \ref{fig:stellarSED}b. 
We find that this can better reproduce both the blue UV slope and the Balmer break, leading to a better overall fit.
We note however that several of our individual photometry measurements yield flatter UV slopes (see Appendix \ref{phot_appendix}), which would favor the single cool star model.
The uncertainties on the photometry ultimately leave room for both fits to be plausible.

Interestingly, our best fit combination of a luminous cool star paired with a hot, similarly luminous companion would be a somewhat unusual evolutionary scenario.
Typically, one might expect the more evolved, cooler star to be the more massive and more luminous object.
We note two important caveats here.
First, the stellar parameters are not fully constrained in this fit due to the number of free parameters being high relative to the number of photometric data points.
Additionally, the two-star fit presented here assumes that both stars are at the same magnification.
In a true lensed two-star system, each component of the binary would travel on a slightly different path through the lens, resulting in different magnifications.
The magnification is directly tied to the inferred luminosity, so accounting for variable magnifications could alter the relative bolometric luminosities of the two stars.
This introduces further degeneracies with the already broad stellar parameter space.
We therefore defer extensive simulations of this scenario for future work.
Spectroscopic observations with \JWST\ NIRSpec, expected in December 2022, can help to address these discrepancies and further constrain the temperature and luminosity of the star/stars.

\section{Conclusions}
\label{sec:conclusions}

We present recently obtained \JWST\ imaging of the $\zphot = 6.2$ gravitationally lensed stellar source Earendel. 
The increased depth and wavelength coverage of these images, combined with the higher angular resolution of \JWST\ compared to \HST, allow us to improve constraints on the magnification and radius of Earendel, further supporting the interpretation of it as a distant lensed star.
We further conclude that, if the light of Earendel is dominated by a single star, that star likely has an effective temperature of 13000--16000 K, indicating that it is likely a B-type giant similar to other lensed stars, or perhaps an LBV star. 
The apparent discrepancies between our best-fit single star model and the observed photometric data allow room for the consideration of multi-star models.
In particular, we note that a two star system with one hot ($T_{{\rm eff}} \sim 34000$ K) and one cooler ($T_{{\rm eff}} \sim 9000$ K) star could produce a better fit to the observed data, though the wide parameter space in this case allows for many similarly well-matched solutions.
These initial photometric constraints, while themselves inconclusive, provide an important guide for our upcoming spectroscopic observations.

\JWST\ NIRSpec observations are due to be carried out in December 2022 under GO 2282, which will shed additional light on this remarkable object.

\section*{Acknowledgements}

This work is based on observations made with the NASA/ESA/CSA James Webb Space Telescope (JWST). The data were obtained from the Mikulski Archive for Space Telescopes (MAST) at the Space Telescope Science Institute (STScI), which is operated by the Association of Universities for Research in Astronomy (AURA), Inc., under NASA contract NAS 5-03127 for \JWST. These observations are associated with program JWST GO 2282.

EZ acknowledges support from the Swedish National Space Board.
MB acknowledges support from the Slovenian national research agency ARRS through grant N1-0238. 
AZ acknowledges support by Grant No. 2020750 from the United States-Israel Binational Science Foundation (BSF) and Grant No. 2109066 from the United States National Science Foundation (NSF), and by the Ministry of Science \& Technology, Israel.
MT acknowledges support by the Australian Research Council Centre of Excellence for All Sky Astrophysics in 3 Dimensions (ASTRO 3D), through project number CE170100013. The Cosmic Dawn Center (DAWN) is funded by the Danish National Research Foundation under grant No. 140.
J.M.D. acknowledges the support of projects PGC2018-101814-B-100 and MDM-2017-0765. 
Y.J-T acknowledges financial support from the European Union’s Horizon 2020 research and innovation programme under the Marie Skłodowska-Curie grant agreement No 898633, the MSCA IF Extensions Program of the Spanish National Research Council (CSIC), and the State Agency for Research of the Spanish MCIU through the Center of Excellence Severo Ochoa award to the Instituto de Astrofísica de Andalucía (SEV-2017-0709)
PAO acknowledges support by the Swiss National Science Foundation through project grant 200020\_207349.

The data presented in this paper were obtained from the Mikulski Archive for Space Telescopes (MAST) at the Space Telescope Science Institute. The specific observations analyzed can be accessed via \dataset[10.17909/2x4w-pd04]{https://doi.org/10.17909/2x4w-pd04}.

\facilities{JWST(NIRCam), HST(WFC3)}

\software{
grizli \citep{Brammer21_grizli},
astropy \citep{astropy:2013,astropy:2018},
photutils \citep{photutils_1.5},
Source Extractor \citep{sextractor},
EAZY \citep{Brammer08_eazy}
}

\appendix

\section{Earendel photometry} \label{phot_appendix}

\begin{figure*}
    \centering
    \includegraphics[width=0.45\textwidth]{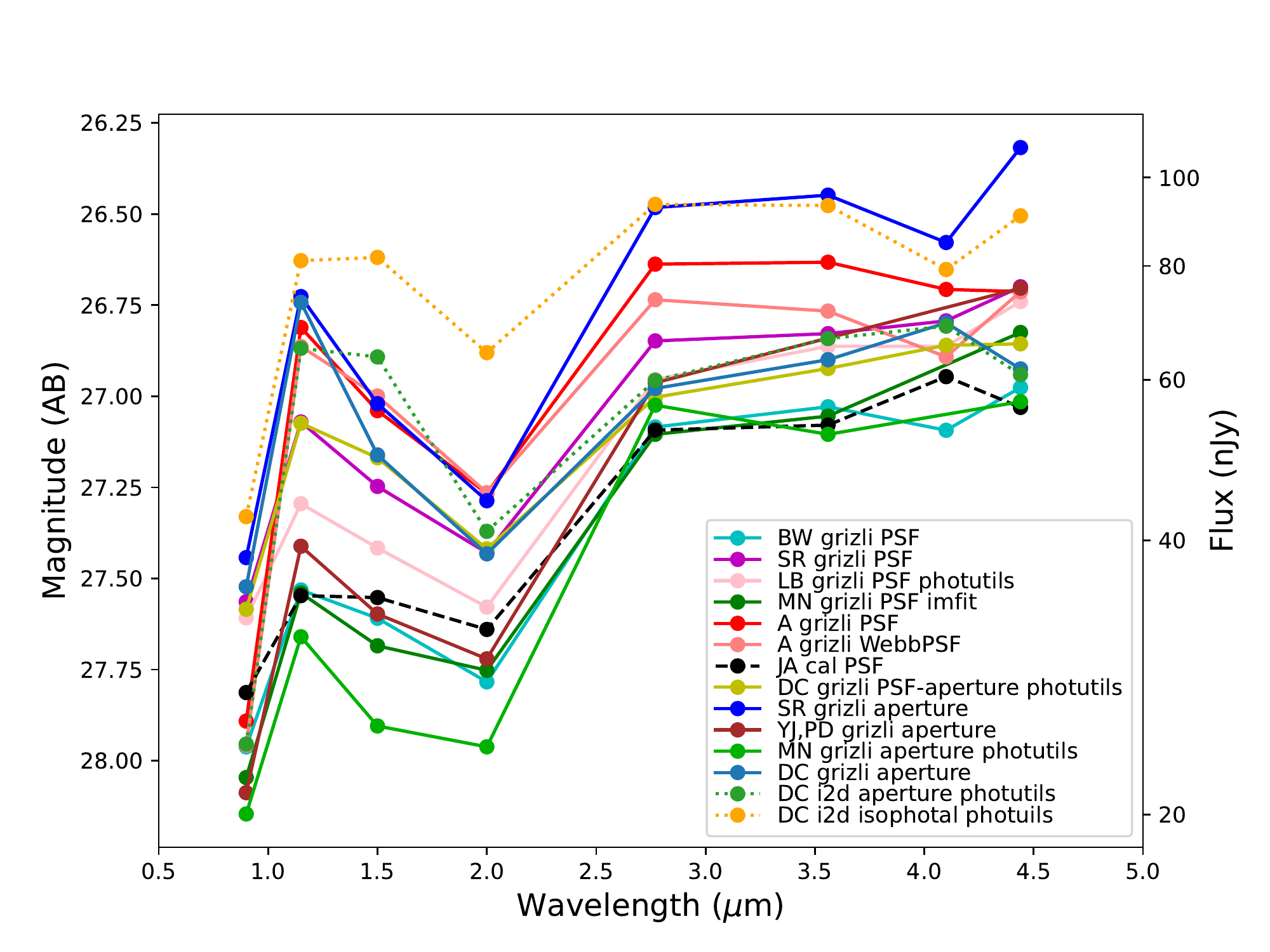}
    \includegraphics[width=0.45\textwidth]{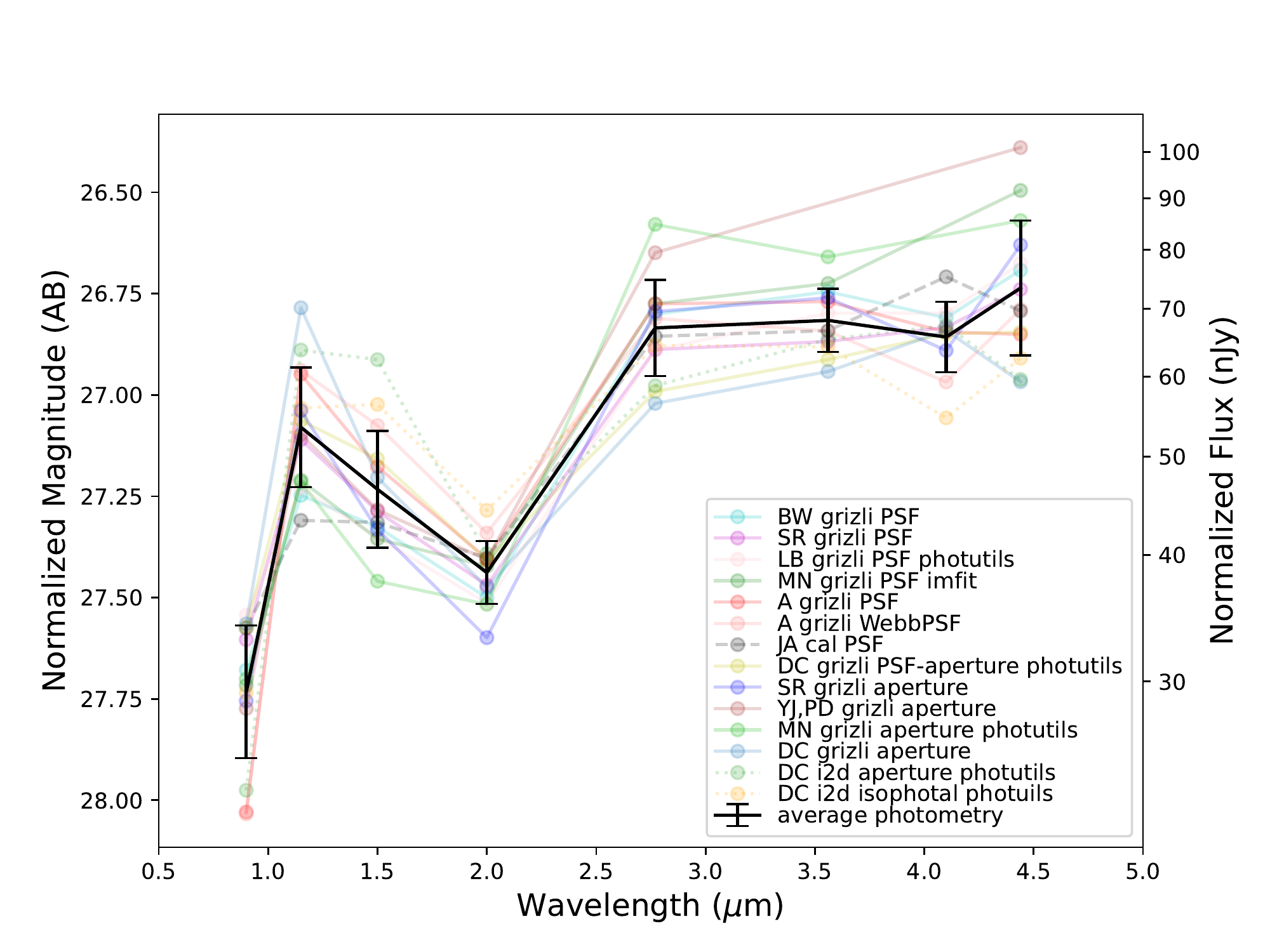}
    \caption{
    {\it Left:} Earendel photometry measurements from 14 analyses by 10 co-authors.
    {\it Right:} Photometry renormalized to an average AB mag 27.1 (across filters)
    along with average SED (across methods) and scatter (errorbars) plotted in black.
    }
    \label{fig:photometry_average}
\end{figure*}

Measurements of Earendel's photometry are complicated by its faint magnitude and location within the curved Sunrise Arc, which makes it difficult to measure and subtract the ``background''.
To mitigate the systematic uncertainties, 
we performed 14 different analyses by 10 co-authors using various methods
described below (Table \ref{tab:photometry_results}).
We then averaged these results to arrive at a
concordance photometry for Earendel in the 8 \JWST\ filters
(Figure \ref{fig:photometry_average}).
A similar approach of averaging photometric redshift results from various methods was shown to be most accurate by \cite{Dahlen13}.
A diversity of perspectives and approaches can improve performance in many fields, also known as the ``wisdom of crowds'' (Surowiecki 2004).

Most of the 14 analyses adopted either aperture photometry or PSF fitting.
In both cases, corrections were made to total flux by accounting for encircled energy within a given aperture size and 
filter.\footnote{\href{https://jwst-docs.stsci.edu/jwst-near-infrared-camera/nircam-performance/nircam-point-spread-functions
}{JDox: NIRCam Point Spread Functions}}
We found that the encircled energies reported in JDox (based on pre-launch estimates)
were consistent with an empirical PSF derived 
from the grizli image reductions based on 4 isolated unsaturated stars.
We used this empirical PSF for most of the PSF fitting analyses.

Analyst A performed PSF-fitting analyses both with this empirical PSF and with WebbPSF models, 
finding very similar results both ways
using the piXedfit software \citep{Abdurrouf21,piXedfit}.

JA independently derived empirical spatially variable PSFs from individual {\tt cal} exposures of the LMC calibration field taken in the various filters (see Anderson et al. 2022, in prep) and used these PSF models to fit Earendel as a point source in the individual {\tt cal} exposures of GO-2282.

SR measured PSF photometry using the DAOPHOT software \citep{Stetson1987_daophot}.
The residual images were inspected and were found to be consistent with the noise level, supporting the interpretation that Earendel is unresolved.
SR also performed aperture photometry using aperture of r=0.2" and applied aperture corrections appropriate for each filter

MN performed photometry in two ways. 
The first, labeled ``aperture" in Table \ref{tab:photometry_results}, uses \texttt{photutils} \citep{photutils} to measure background-subtracted flux in an variable aperture large enough to encircle 90\% of the empirical PSF. 
The second method, labeled ``PSF" in Table \ref{tab:photometry_results}, utilizes \texttt{imfit} \citep{Erwin15_imfit} to make a model of Earendel in each filter.
The total flux is then measured in the same apertures as the ``aperture" method.

BW fit point source models convolved with the grizli empirical PSFs to each filter using a custom code similar to the forward model described in \citet{Welch22_clumps}.

DC performed a hybrid analysis cloning the empirical PSF at two locations on the Sunrise Arc on either side of Earendel $0.6''$ away.
They derived the photometry within a $r = 0.2''$ (SW) or $0.4''$ (LW)
aperture that best matched
Earendel's photometry in the same aperture.
The two locations sampled different background levels, 
and the results were averaged.

DC also measured aperture photometry in $r = 0.2''$ apertures
(finding consistent results for $0.2''$ and $0.3''$ 
after applying encircled energy aperture corrections).
They found flux measurements varied 10--20\% depending on where they measured the background: on either side of Earendel $0.6''$ away along the Sunrise Arc. 
They averaged measurements from the NE and SW sides.

YJ \& PD identified the pixels associated to Earendel using NoiseChisel \citep{akh15} clumps map, disabling the kernel option previously, and interpolated the background in this region based on flux from all surrounding pixels. This ``background" naturally folds the contribution from the sky, the Sunrise Arc, the wings of the nearby galaxies, and the intracluster light. Earendel's flux is obtained as the difference between the original and the interpolated images, thus minimizing the impact of the uncertainty in the sky on the final measurement. An aperture of $r = 0.3''$ was used.

DC extracted results from photutils analysis of the full {\tt i2d} images 
(all aligned to the F200W image pixels),
with objects detected in the F200W image, 
and F200W PSF-matched to each LW filter to measure the LW colors
(without PSF corrections for the SW colors).
Photometry was measured within both round and isophotal apertures, 
subtracting backgrounds measured in annuli around Earendel.


Finally, for each analysis, we calculate the average magnitude across all filters.
This varied by 0.9 mag across all methods,
reflecting variations in the total flux normalization.
We took the average of these normalizations, AB mag 27.1,
and renormalized all SEDs to this average across filters.
This corrects for variations in total flux measurements without altering the SED derived by each method.
We then calculate the average and scatter (RMS) across all methods
as the final magnitude and uncertainty for Earendel in each filter.
These results are plotted in Figure \ref{fig:photometry_average}.

We note that 
all analyses derive a red F200W$-$F277W color 
(Balmer excess indicative of a cooler star $T \sim 10000$ K)
and almost all derive a blue F115W$-$F200W color 
(rest-UV slope indicative of a hotter star $T \sim 30000$ K).
The JA photometry yields a flat rest-UV slope
that could be well fit by a single star
(see e.g., Figure \ref{fig:stellarSED}).

We also tried restricting our analysis to the PSF photometry analyses.
The resulting SED is consistent within the uncertainties
(to that shown in \ref{fig:photometry_average})
with a similar Balmer break and slightly shallower rest-UV slope.




\begin{table}[]
    \centering
    \begin{tabular}{c c c c c c c c c c c c}
        Analyst & images & method & software & F090W & F115W & F150W & F200W & F277W & F356W & F410M & F444W\\
        initials &  &  &  & nJy & nJy  & nJy & nJy  & nJy & nJy  & nJy & nJy\\
         \hline
         
BW & {\tt grizli} & PSF &  & 23 & 35 & 32 & 28 & 53 & 56 & 52 & 58  \\ 
SR & {\tt grizli} & PSF & DAOPHOT & $34 \pm 9$ & $54 \pm 9$ & $45 \pm 8$ & $38 \pm 7$ & $66 \pm 11$ & $67 \pm 8$ & $69 \pm 10$ & $76 \pm 7$  \\ 
LB & {\tt grizli} & PSF & photutils & $32 \pm 2$ & $43 \pm 2$ & $39 \pm 2$ & $33 \pm 1$ & $60 \pm 3$ & $65 \pm 2$ & $65 \pm 3$ & $72 \pm 2$  \\ 
MN & {\tt grizli} & PSF & imfit & 21 & 35 & 30 & 28 & 52 & 54 & \nodata & 67  \\ 
A & {\tt grizli} & PSF & piXedfit & 25 & 68 & 55 & 45 & 80 & 80 & 75 & 75  \\ 
A & {\tt grizli} & WebbPSF & piXedfit & 23 & 65 & 57 & 45 & 73 & 71 & 63 & 75  \\ 
JA & {\tt cal} & PSF &  & $27 \pm 1$ & $35 \pm 4$ & $34 \pm 3$ & $31 \pm 3$ & $52 \pm 9$ & $53 \pm 4$ & $60 \pm 2$ & $56 \pm 2$  \\ 
DC & {\tt grizli} & PSF-aperture & photutils & $33 \pm 2$ & $54 \pm 2$ & $49 \pm 5$ & $38 \pm 3$ & $57 \pm 3$ & $61 \pm 9$ & $65 \pm 0$ & $65 \pm 14$  \\ 
SR & {\tt grizli} & aperture &  & $38 \pm 32$ & $74 \pm 39$ & $56 \pm 34$ & $44 \pm 32$ & $92 \pm 39$ & $95 \pm 40$ & $84 \pm 39$ & $107 \pm 41$  \\ 
YJ,PD & {\tt grizli} & aperture & NoiseChisel & $21 \pm 7$ & $39 \pm 7$ & $33 \pm 5$ & $29 \pm 6$ & $59 \pm 7$ & \nodata & \nodata & $75 \pm 10$  \\ 
MN & {\tt grizli} & aperture & photutils & 20 & 31 & 25 & 24 & 56 & 52 & \nodata & 56  \\ 
DC & {\tt grizli} & aperture & photutils & 35 & 73 & 49 & 38 & 58 & 63 & 69 & 61  \\ 
DC & {\tt i2d} & aperture & photutils & 23 & 64 & 63 & 40 & 59 & 66 & 68 & 60  \\ 
DC & {\tt i2d} & isophotal & photutils & 42 & 81 & 81 & 64 & 93 & 93 & 79 & 90  \\ 
\hline
    \end{tabular}
    \caption{\JWST\ NIRCam photometry of Earendel from 14 analyses.
    Most performed photometry using the grizli image reductions. 
    Some used the pipeline products directly: 
    either the {\tt cal} or {\tt i2d} images.
    }
\label{tab:photometry_results}
\end{table}

\bibliography{masterbib.bib}

\end{document}